\documentclass[%
prl,
 reprint,
superscriptaddress,
preprintnumbers,
nofootinbib,
 amsmath,amssymb,
 aps,
]{revtex4-1}
\pdfoutput=1
\usepackage{hyperref}

\usepackage[textwidth=1.25cm,textsize=footnotesize,
]{todonotes}

\usepackage[utf8]{inputenc}

\newcommand{\be}{\begin{equation}}
\newcommand{\ee}{\end{equation}}
\newcommand{\ba}{\begin{eqnarray}}
\newcommand{\ea}{\end{eqnarray}}
\renewcommand{\d}{\partial}
\renewcommand{\l}{\left(}
\renewcommand{\r}{\right)}

\newcommand{\e}{\mathrm{e}}

\makeatletter
\def\gsim{\compoundrel>\over\sim}
\def\lsim{\compoundrel<\over\sim}
\def\compoundrel#1\over#2{\mathpalette\compoundreL{{#1}\over{#2}}}
\def\compoundreL#1#2{\compoundREL#1#2}
\def\compoundREL#1#2\over#3{\mathrel
         {\vcenter{\hbox{$\m@th\buildrel{#1#2}\over{#1#3}$}}}}
\makeatother

\begin{document}

\preprint{INR-TH/2018-009}

\title{Constraints on hidden photons produced in nuclear reactors}
\author{Mikhail Danilov}
\email{mvdanil@gmail.com}
\affiliation{Lebedev Physical Institute of the Russian Academy of Sciences,
  Moscow 119991, Russia}

\author{Sergey Demidov}
\email{demidov@ms2.inr.ac.ru}
\affiliation{Institute for Nuclear Research of the Russian Academy of Sciences,
  Moscow 117312, Russia}
\affiliation{Moscow Institute of Physics and Technology, 
  Dolgoprudny 141700, Russia}

\author{Dmitry Gorbunov}
\email{gorby@ms2.inr.ac.ru}
\affiliation{Institute for Nuclear Research of the Russian Academy of Sciences,
  Moscow 117312, Russia}
\affiliation{Moscow Institute of Physics and Technology, 
  Dolgoprudny 141700, Russia}

\begin{abstract}
New light vector particles -- hidden photons -- are present in many
extensions of the Standard Model of particle physics. They can be
produced in nuclear reactors and registered by neutrino detectors. We
obtain new limits on the models with the hidden photons 
from an analysis of published results of the 
TEXONO neutrino experiment.
Accounting for
oscillations between the visible and hidden photons, we find that the
neutrino experiments are generally insensitive to the hidden photons 
lighter than $\sim 0.1$\,eV.
\end{abstract}

\maketitle


A number of extensions of the Standard Model of particle physics (SM)
introduce massive vectors, singlet with respect to the SM gauge
group. These hypothetical particles are called hidden photons, dark
photons, paraphotons \cite{Okun:1982xi}. They can serve as dark
matter particles or messengers between the visible sector (SM) and
hidden sector(s), which dynamics solves (some of) SM phenomenological
and theoretical problems (neutrino oscillations, matter-antimatter
asymmetry of the Universe, etc) or suggests some phenomena which
impacts on physics and the Universe are not recognizable at
present (see e.g. \cite{Pospelov:2007mp,Bjorken:2009mm,Ilten:2018crw}).  

The hidden photon can couple to the SM via vector portal
interaction. The 
corresponding coupling constant  is dimensionless and hence low- and
high-energy experiments exhibit similar sensitivity to this type of
new physics, if the hidden photon is sufficiently light. In
particular, the hidden photon $X_\mu$ of mass $m_X$ can mix with
the SM photon
$A_\mu$, the relevant lagrangian~\cite{Holdom:1985ag} reads
\begin{equation}
  \label{lagr}
        {\cal L} = - \frac{1}{4}\, F_{\mu\nu}^2 - \frac{1}{4}\, X_{\mu\nu}^2
        - \frac{\epsilon}{2}X_{\mu\nu}F^{\mu\nu} + \frac{m_X^2}{2}
        X_\mu^2 - e A_\mu j^\mu_{em}\,.
\end{equation}
Here $F_{\mu\nu}\equiv \d_\mu A_\nu-\d_\nu A_\mu$, $X_{\mu\nu}\equiv
\d_\mu X_\nu-\d_\nu X_\mu$; $\epsilon$ is a dimensionless parameter of
visible-hidden photon mixing and $j^\mu_{em}$ is electromagnetic current.

The mixing term in \eqref{lagr} can be responsible for the hidden photon
production in a nuclear reactor, if kinematically allowed, hidden
photons of 
i.e. $m_X\!\lesssim\! 1$\,MeV can be produced by photons of energy
$E_\gamma\sim1\!-\!10$\,MeV. 

For sufficiently small $m_X$ the mixing in~\eqref{lagr} converts
photons into hidden photons via oscillations similar to the well-studied
neutrino oscillations. The visible-to-hidden photon oscillations are
fully described e.g. in Ref.\,\cite{Redondo:2015iea} devoted to the
hidden photon production and propagation in the Sun. Given the above
analogy with neutrino oscillations, we consider the interaction
eigenstates to be more  convenient for the estimate of hidden photon
yield. Typically, one replaces (see, e.g.~\cite{Redondo:2008aa}) the
hidden photon in \eqref{lagr} as $X_\mu\to S_\mu  - \epsilon A_\mu$,
with $S_\mu$ being sterile with respect to the electromagnetic
interaction. The kinetic term of \eqref{lagr} becomes diagonal in
terms of the new variables, while the mass term gains off-diagonal
components. The interaction with the electromagnetic current produces a
quantum wave packet of photon $A_\mu$, which is not a propagation
eigenstate. In the relativistic regime the evolution of transverse modes of
the photon--hidden photon system can be described by the
following Hamiltonian (see, e.g.~\cite{Redondo:2013lna})
  \begin{equation}
    \label{hamilt}
    H = \frac{1}{2E_\gamma}\left(\begin{array}{cc}
      \epsilon^2 m_X^2 + m_\gamma^2 & -\epsilon m_X^2 \\
      -\epsilon m_X^2 & m_X^2
      \end{array}\right)\;.
  \end{equation}
Here $m_\gamma$ is an effective mass of photon which the latter
acquires due to its coherent forward scattering off free electrons of
the media~\cite{Braaten:1993jw}. The value of $m_\gamma$ coincides
with the corresponding plasma frequency~\cite{ll10},
$m_\gamma^2=4\pi\alpha n_e/m_e$, where $\alpha$ is fine-structure
constant, $m_e$ is electron mass and $n_e$ is density of free
electrons
\footnote{For the reactor photon energies all electrons in
  matter can be considered as free.}
inside the 
reactor. Numerically, for the reactor material one finds that
the photon mass varies within 
\begin{equation}
  \label{photon-mass}
  m_\gamma\simeq 20-60\,\text{\rm eV}\,.
\end{equation}
The Hamiltonian~\eqref{hamilt} determines propagation eigenstates of
the system which are mixtures of $A_\mu$ and $S_\mu$.
In terms of interaction eigenstates $S_\mu$ and $A_\mu$, their evolution
looks as oscillations, with the transition probability
  $P\l\gamma\to\gamma'\r =4\epsilon^2\sin^2{\frac{\Delta
      m^2L}{4E_\gamma}}$ valid for negligible absorption. 
The corresponding oscillation length is given by
\begin{equation}
\label{osc-length}
  L_{osc}\approx 2.5\,\text{\rm cm}
\times\frac{E_\gamma}{1\,\text{\rm MeV}}\frac{\l 10\,\text{\rm eV}\r^2}{\Delta m^2}\,,
\end{equation}
where the mass squared difference $\Delta m^2$ reads
\begin{equation}
  \label{mass2diff}
    \Delta m^2= \sqrt{\l m_X^2-m_\gamma^2\r^2+2\epsilon^2 m_X^2(m_X^2+m_\gamma^2)}\,.
\end{equation}
The latter never falls below $m^2_\gamma\gtrsim (20\,\text{\rm eV})^2$, except for
the region, $m_X\approx m_\gamma$,  where the oscillation
length \eqref{osc-length} is largest, and which we call the
resonance 
region in what follows. Hence, the oscillation length
\eqref{osc-length} is typically much smaller than the size of the
nuclear reactor core of order meters. 

The oscillations between the states $S_\mu$ and $A_\mu$
proceed until corresponding wave packets of the Hamiltonian
eigenstates get separated in space due to their different velocities,
or photon interacts in media; both 
processes naturally break the quantum coherence between the two
oscillating states and invalidate the oscillation approximation to the
hidden photon production. The relevant coherence length due to
space separation of wave packets in vacuum can be estimated as
$l_{coh}\sim \sigma / \Delta v$ where $\sigma$ is size 
of the photon wave packet at production and $\Delta v$ is difference
in velocities of the mass eigenstates. Using typical half-lifes of
prompt $\gamma$-decays of fission fragments about
$10^{-12}-10^{-11}$\,s (see
e.g.~\cite{JOHANSSON1964378,1402-4896-3-3-4-005}),   
one obtains $\sigma\sim 0.03-0.3$\,cm. The corresponding
coherence length   
$
l_{coh} \sim 6\cdot 10^{8-9}\,\text{\rm cm}\,\times
\l E_\gamma/1\,\text{\rm MeV}\r^2
\l 10\,\text{\rm eV}\r^2/\Delta m^2 
$
always exceeds the oscillation length \eqref{osc-length}.
This statement remains true even in the media where
the coherence length decreases due to interactions of fission
products and after rescattering of photons off electrons reducing
$\sigma$ to $10^{-6}$~cm. So the oscillation approximation is
justified in our case.  

The oscillation is terminated by absorption of the photon in the reactor
material. The photon absorption length in a nuclear
reactor, $1/\Gamma$, varies from a few to few tens cm for the 
energy range
$1-10$\,MeV.
Its effect can be
  described by the replacement $\frac{m^2_\gamma}{2E_\gamma}
  \to \frac{m^2_\gamma}{2E_\gamma} - i\Gamma$ in the
  Hamiltonian~\eqref{hamilt}. 
The absorption length is much  shorter than the
reactor size of about several meters, but it is typically longer than
the oscillation length \eqref{osc-length} except for the resonance
region $\Delta m^2\approx 0$. In the following numerical estimates we
take $1/\Gamma=10$\,cm. Most photons produced in the core get
absorbed in the reactor material (mostly in water and steel) unless
they oscillate into hidden photons with the
probability\,\cite{Redondo:2015iea}
\begin{equation}
  \label{probability}
  P=\epsilon^2\times\frac{m_X^4}{\l
 \Delta m^2\r^2+ E_\gamma^2 \Gamma^2}\,,  
\end{equation}
which can be obtained straightforwardly by solving the Schrodinger
equation with the Hamiltonian~\eqref{hamilt} for distances larger than the absorption
length. The term in denominator
of~\eqref{probability} responsible for the photon absorption dominates
when  
\begin{equation}
  \label{condition}
\Delta m^2 \ll E_\gamma \Gamma \approx
2\times\l\frac{E_\gamma}{1~{\rm
    MeV}}\r\l\frac{10\,\text{\rm cm}}{1/\Gamma}\r~{\rm eV}^2\,,  
\end{equation} 
i.e. in the resonance region, where $m_X\approx m_\gamma$ with
the accuracy of a few percent. 

In the non-resonance case the condition \eqref{condition}
 is opposite, and the probability depends on the relation between
$m_X$ and $m_\gamma$. For heavier hidden photons, i.e. when $m_X\gg
m_\gamma>20$\,eV, the probability turns to simple $P=\epsilon^2$ law.
In the absence of coherence in $\gamma$-$X$ system this result follows 
from the calculation of Compton scattering with $X$ emerging due to
mixing with photon. (This result deviates from that in
Ref.\,\cite{Park:2017prx}, where a numerical factor $2/3$ was
introduced to account for difference in numbers of polarization states
between photon (two) and massive photon (three). We find this factor
irrelevant since only two transverse polarizations of the hidden
photon are produced via oscillations of the massless photon. The
production of the longitudinal component of the massive photon is
suppressed for 
the masses and photon energies of interest, $P\propto
m_X^2/E_\gamma^2$, see e.g.~\cite{An:2013yfc,Redondo:2013lna}.)  
In the opposite case $m_X\ll m_\gamma$ one obtains from
  Eq.~\eqref{probability}
\begin{equation}
P\approx 6\times 10^{-6}\times \epsilon^2\times \left(\frac{m_X}{1~{\rm
    eV}}\right)^4,   
\end{equation}
where we use $m_\gamma=20$\,eV for the estimate.

The hidden photon production rate is obtained by convolution of the
probability \eqref{probability} with the photon flux in the reactor, which
we take normalized to that measured~\cite{FRJ} 
for $E_\gamma\gtrsim 0.2$\,MeV from the FRJ-1 reactor core
\[
\frac{dN_\gamma}{dE_\gamma}=0.58\times 10^{21}\times \frac{T}{\text{\rm GW}}
\times \e^{-\frac{E_\gamma}{0.91\,\text{\rm MeV}}}\,
\]
in units photons/(s$\times$MeV),
with $T$ being the reactor thermal power. 
The hidden photons leave the
reactor and can be observed at some distance in a detector designed to
measure the reactor antineutrino flux. The hidden photons oscillate and produce
photons which can be detected
  via the Compton
scattering off electrons. 
Each hidden photon then produces the Compton-like signature with
the probability 
\begin{equation}
  \label{probability_det}
  P=\epsilon^2\times\frac{m_X^4}{\l
    \Delta m^2\r^2}\,,
\end{equation}
(see~\eqref{probability}, where we set  $\Gamma=0$, which can be done away from the
  resonance region) 
and $m_\gamma$ in $\Delta m^2$ \eqref{mass2diff} should be calculated
for material along the hidden 
photon path inside the detector. Even 
neglecting $m_\gamma$ (that is for $m_X\gg m_\gamma$) one obtains an
estimate $P\approx \epsilon^2$.

However, the neutrino detectors are made of dense material, so the
effective photon mass is certainly not smaller than that in water. For
the numerical estimates below we chose $m_\gamma=20$\,eV
inside neutrino detector as well. Hence, the total probability in
general non-resonance case is just a product of \eqref{probability}
and \eqref{probability_det}, which implies a huge suppression factor
for the light hidden photons, $m_X\ll m_\gamma$.

In the resonance case, $m_\gamma=m_X$, the mass difference \eqref{mass2diff} becomes equal
to 
\[
\Delta m^2=0.8 \times \frac{\epsilon}{10^{-3}} \l \frac{m_\gamma}{20\,\text{\rm eV}}\r^2\,\text{\rm eV}^2\,,
\]
and the oscillation length \eqref{osc-length} is
\[
L_{osc}=3\,\text{\rm m}\,\frac{E_\gamma}{1\,\text{\rm MeV}}
\frac{10^{-3}}{\epsilon}
\l \frac{20\,\text{\rm eV}}{m_\gamma}\r^2\,. 
\]
One observes, that for $\epsilon<10^{-3}$ condition \eqref{condition}
is obeyed and the probability \eqref{probability} 
reduces  to
\begin{equation}
\label{P_amplified}
P \approx 4\times 10^{4}\times \epsilon^2\times
\left(\frac{m_\gamma}{20\,\text{\rm eV}}\right)^4\times\left(\frac{1~{\rm
    MeV}}{E_\gamma}\right)^2\,.
\end{equation}
One expects similar enhancement for the
  hidden-to-visible conversion probability, however it
generally occurs for another resonance mass $m_X$, since reactor and
detector materials are different, and so the corresponding photon
masses. 

Therefore, in the two very narrow mass ranges the number of signal
events in the detector gets amplified hundred thousand times with respect to
the Compton-based result, $P=\epsilon^2$. Since the estimate \eqref{P_amplified} is
valid only in the very narrow mass ranges defined by the photon
effective mass in the matter, its applicability requires a good
knowledge of the nuclear reactor core structure and the detector, which is unavailable
for us. However, it can be applied by e.g. the NEOS and TEXONO
collaborations. Note, that in the realistic case of inhomogeneous materials the
probability formula \eqref{probability} gets modified,
see Ref.~\cite{Redondo:2015iea}.


A side remark concerns the recent paper\,\cite{Park:2017prx}
where a study similar to ours was
performed but the oscillations were neglected. We find 
that at $m_X\gtrsim 20$\,eV its results for the event numbers are
underestimated by factor 3/2, except the resonance regions where they
may be (a special study is needed) underestimated by a factor of
$\sim 10^5$. For lighter hidden photons, $m_X\lesssim 20$\,eV, the number of
signal events are overestimated by factor $2/3\times
(m_\gamma/m_X)^8$, see expressions for the conversion
  probabilities~\eqref{probability} and~\eqref{probability_det}. While Ref.\,\cite{Park:2017prx} claims the
mass-independent upper limit of about $\epsilon\lesssim 10^{-5}$, our
observation suggests, instead, that the neutrino experiments are
absolutely insensitive to the hidden photons lighter than about
0.05\,eV.

Now we turn to the analysis of the experimental data of the 
TEXONO neutrino experiment~\cite{Deniz:2009mu}, and use its result to
place limits
\footnote{See Ref.~\cite{Lindner:2018kjo} for similar   study in a model where 
   new massive vector boson directly couples to neutrinos.}
on mixing $\epsilon$. 
%
To measure the $\bar\nu_e-e^-$ scattering cross section the TEXONO
experiment used scintillator crystal detector, located at 28 m from
the core of $T=2.9$\,GW thermal-power reactor. With electron recoil
energy in 3-8\,MeV range the TEXONO collaboration extracted
$414\pm100.6$ events in 160 days.  This number is $30.7\pm100.6$
events larger than the SM expectation.  The excess is smaller than
195.7 events at the 95\% CL.  We use this number to determine the 95\%
CL upper limit on the number of hidden photons detected in the TEXONO
experiment. 
The TEXONO collaboration applies a special anti-Compton
selection, which reduces the background by a factor of 6 in the
energy range 3-8\,MeV utilized for the searches. The signature of a
hidden photon is identical to the Compton scattering
process. Therefore, the anti-Compton selection decreases the
efficiency of the hidden photon detection. The suppression factor of
the anti-Compton selection for single photons is not given by the
TEXONO collaboration. It can be smaller than that for the background
suppression. However, for the upper limit estimates we can
assume them to be equal. This leads to the upper limit of 1174.1
hidden photon events in the TEXONO experiment at the 95\% CL.

\begin{figure}[htb]
  \centerline{
    \includegraphics[width=1.0\columnwidth]{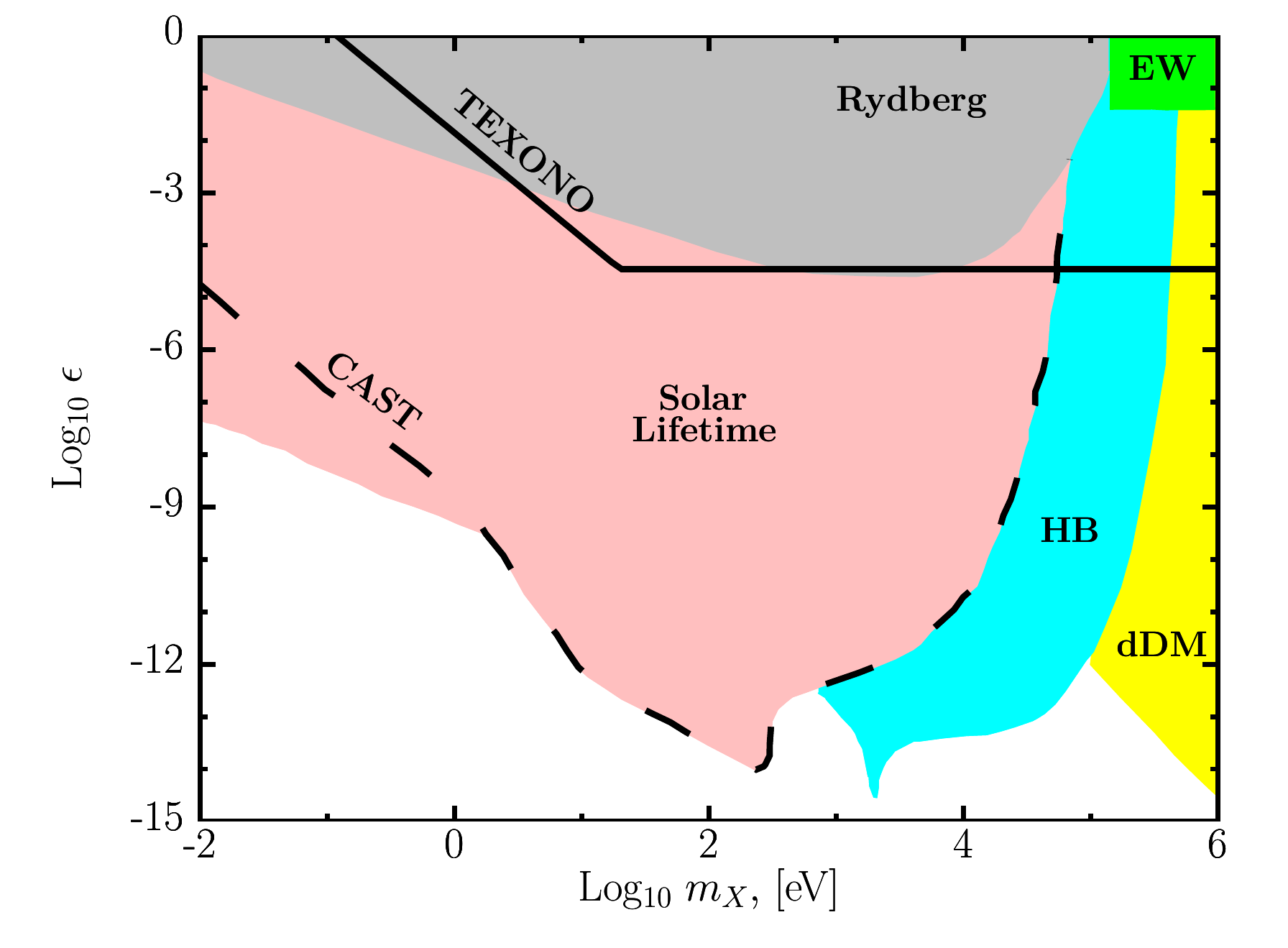}}
  \caption{\label{limits} 95\% CL exclusion upper limit in the
    parameter plane $(m_X, \epsilon)$ of the hidden photon model from
    TEXONO experiment (black solid line) in comparison with other
    results (see Ref.~\cite{Jaeckel:2010ni} for details). }
\end{figure}

In Fig.~\ref{limits} we present a revised 95\% CL upper limit on the
parameters $\epsilon$ and $m_X$ of the model from TEXONO data where in
recalculation of the results from\,\cite{Park:2017prx} we take into
account both the theoretical and experimental issues discussed above.
At large masses of the hidden photon, i.e. $m_X\gsim 20$~eV,
the overall correction to the signal estimate claimed 
in~\cite{Park:2017prx} is a factor of 4; given the $\epsilon^4$
dependence the corresponding limit on $\epsilon$ is only slightly
weaker than  
that presented in~\cite{Park:2017prx}. At the same time we see
that the sensitivity of the reactor experiment to the hidden 
photon model is drastically decreased for $m_X\lsim 20$~eV.

A side remark concerns  Ref.\,\cite{Park:2017prx}, where TEXONO and
NEOS data were both analyzed to constrain $\epsilon$.
For the TEXONO data analysis in \cite{Park:2017prx} the suppression of
the hidden photon detection efficiency due to the anti-Compton
selection was ignored. This underestimates the 
upper limit on the number of hidden photons by a factor which can be
as large as 6. Analysis of the NEOS data is wrong at several points. Since Ref.\,[23]
from\,\cite{Park:2017prx} is not available for us, we naturally
assume that the conclusion 'all of the reactor-on event candidates are
due to background' is based on the approximate equality
$N_{on}/T_{on}=N_{off}/T_{off}$. Here $N_{on}$ ($N_{off}$) and
$T_{on}$ ($T_{off}$) are the number of $e/\gamma$-events and data
taking time during the reactor on (off) period. Then the absence of the signal events
associated with the hidden photons implies
\begin{equation}
\label{S-events}
 N_{hph}=N_{on}-N_{off}\,\frac{T_{on}}{T_{off}}\approx 0\,. 
\end{equation}
Its statistical uncertainty 
\[
\Delta
N_{hph}=\sqrt{N_{on}+N_{off}\,\frac{T^2_{on}}{T^2_{off}}}=\sqrt{N_{on}\l
  1+ \frac{T_{on}}{T_{off}}\r} 
\]
determines the upper limit on the hidden-visible mixing, 1.64$\Delta
N_{hph}$ would correspond to 95\% CL. It is worth to mention that
the two-sided CL were used in \cite{Park:2017prx} instead of one-sided.
Therefore, the obtained upper limits correspond to 97.5\%CL instead of
95\% CL. Numerically
for the NEOS on- and off- time intervals we obtain $\Delta
N_{php}\approx 2.2\times\sqrt{N_{on}}$ instead of $\sqrt{N_{on}}$
adopted in\,\cite{Park:2017prx}.  

Even more important is that the estimation of the upper limit ignores
possible systematic errors. The upper limit corresponds to $7.3\times
10^{-5}$ of the total number of $e/\gamma$-events. This small number
implicitly assumes the absence of time variations of the detector
efficiency and background contribution at a similar very low relative
level. The NEOS experiment \cite{Ko:2016owz} has not presented any
evidence of such a challenging stability.

Indeed, signatures of the hidden photon interactions in the NEOS
detector are practically indistinguishable from signatures of
positrons in the electron antineutrino induced Inverse Beta Decay
(IBD) reactions. The NEOS experiment detected 339.1 thousand IBD
events whose signature includes apart from a prompt positron signal also
a delayed event resulting from $n$-Gd capture. At least 292.7 thousand
of them (1626 events per day for data taking of 180 days) had the total 
positron energy (prompt energy) in the range 1-5\,MeV, at which the
estimation of the expected number of hidden photons was performed. This
number is 5.5 times larger than the 95\% CL upper limit
on number of observed hidden photon events presented in the
paper\,\cite{Park:2017prx}.
{ Here we again point out that the hidden
photon event candidates would look like single prompt events and thus
their number can not be smaller than the number of fully reconstructed
IBD events divided by the neutron detection efficiency and other IBD
process selection efficiencies.
This observation again explicitly demonstrates that the 
estimation \,\cite{Park:2017prx} of the upper limit on the number of
detected hidden photons in the NEOS experiment can not be correct.
It is at least a factor of 6 too optimistic.}
For a typical stability
of the detector efficiency and background level of about 1\% the limit
on the number of the detected hidden photons in the NEOS experiment
derived in \,\cite{Park:2017prx} is 2 orders of magnitude too
optimistic. That is why we do not use the NEOS data in our calculations
of the upper limit on the mixing parameter $\epsilon$ shown in
Fig.~\ref{limits}.


To conclude, we present theoretical description of the expected
dark photon 
signal in reactor neutrino experiments and obtain corrected upper
limits on the hidden photon mass $m_X$ and the mixing parameter
$\epsilon$ from the TEXONO data, see Fig.~\ref{limits}. There are two
comments in order. First, we
  disregard Compton rescattering of the photons during their
  propagation in the reactor core. The secondary photons may
  contribute to the hidden photon production. In this respect our
  limits are conservative. 
 Second, in our estimates we take
  minimum value of the effective photon mass equal $m_\gamma\approx
  20$\,eV for the reactor and detector. The sensitivity to lighter
  hidden photons, $m_X<m_\gamma$, is strongly reduced. The chosen value refers
  to water and grows in denser materials, thus suppressing the
  sensitivity for correspondingly heavier hidden photons. Therefore, 
we expect, that limits in 
Fig.~\ref{limits} may be  
improved, but only mildly, with dedicated analyses performed by the
NEOS and TEXONO collaborations. The only promising case is the
resonance, $m_X\approx m_\gamma$, where the amplified production of
the hidden photons may significantly improve the limit on mixing
$\epsilon$. 

\vskip 0.3cm
We thank S.\,Gninenko, S.\,Troitsky and I.\,Tkachev for valuable
discussions. 
The theoretical analysis of hidden photon production and detection
within the oscillation approximation was supported by the RSF grant
17-12-01547. The analysis of the experimental data adopted to limit
the model parameters was supported
by the Grant of the Russian
Federation Government, Agreement \#14.W03.31.0026 from 15.02.2018.

\bibliographystyle{apsrev4-1}
\bibliography{HPh}

\begin{thebibliography}{19}%
\makeatletter
\providecommand \@ifxundefined [1]{%
 \@ifx{#1\undefined}
}%
\providecommand \@ifnum [1]{%
 \ifnum #1\expandafter \@firstoftwo
 \else \expandafter \@secondoftwo
 \fi
}%
\providecommand \@ifx [1]{%
 \ifx #1\expandafter \@firstoftwo
 \else \expandafter \@secondoftwo
 \fi
}%
\providecommand \natexlab [1]{#1}%
\providecommand \enquote  [1]{``#1''}%
\providecommand \bibnamefont  [1]{#1}%
\providecommand \bibfnamefont [1]{#1}%
\providecommand \citenamefont [1]{#1}%
\providecommand \href@noop [0]{\@secondoftwo}%
\providecommand \href [0]{\begingroup \@sanitize@url \@href}%
\providecommand \@href[1]{\@@startlink{#1}\@@href}%
\providecommand \@@href[1]{\endgroup#1\@@endlink}%
\providecommand \@sanitize@url [0]{\catcode `\\12\catcode `\$12\catcode
  `\&12\catcode `\#12\catcode `\^12\catcode `\_12\catcode `\%12\relax}%
\providecommand \@@startlink[1]{}%
\providecommand \@@endlink[0]{}%
\providecommand \url  [0]{\begingroup\@sanitize@url \@url }%
\providecommand \@url [1]{\endgroup\@href {#1}{\urlprefix }}%
\providecommand \urlprefix  [0]{URL }%
\providecommand \Eprint [0]{\href }%
\providecommand \doibase [0]{http://dx.doi.org/}%
\providecommand \selectlanguage [0]{\@gobble}%
\providecommand \bibinfo  [0]{\@secondoftwo}%
\providecommand \bibfield  [0]{\@secondoftwo}%
\providecommand \translation [1]{[#1]}%
\providecommand \BibitemOpen [0]{}%
\providecommand \bibitemStop [0]{}%
\providecommand \bibitemNoStop [0]{.\EOS\space}%
\providecommand \EOS [0]{\spacefactor3000\relax}%
\providecommand \BibitemShut  [1]{\csname bibitem#1\endcsname}%
\let\auto@bib@innerbib\@empty
\bibitem [{\citenamefont {Okun}(1982)}]{Okun:1982xi}%
  \BibitemOpen
  \bibfield  {author} {\bibinfo {author} {\bibfnamefont {L.~B.}\ \bibnamefont
  {Okun}},\ }\href@noop {} {\bibfield  {journal} {\bibinfo  {journal} {Sov.
  Phys. J. Exp. Theor. Phys.}\ }\textbf {\bibinfo {volume} {56}},\ \bibinfo
  {pages} {502} (\bibinfo {year} {1982})},\ \bibinfo {note} {[Zh. Eksp. Teor.
  Fiz.83,892(1982)]}\BibitemShut {NoStop}%
\bibitem [{\citenamefont {Pospelov}\ \emph {et~al.}(2008)\citenamefont
  {Pospelov}, \citenamefont {Ritz},\ and\ \citenamefont
  {Voloshin}}]{Pospelov:2007mp}%
  \BibitemOpen
  \bibfield  {author} {\bibinfo {author} {\bibfnamefont {M.}~\bibnamefont
  {Pospelov}}, \bibinfo {author} {\bibfnamefont {A.}~\bibnamefont {Ritz}}, \
  and\ \bibinfo {author} {\bibfnamefont {M.~B.}\ \bibnamefont {Voloshin}},\
  }\href {\doibase 10.1016/j.physletb.2008.02.052} {\bibfield  {journal}
  {\bibinfo  {journal} {Phys. Lett.}\ }\textbf {\bibinfo {volume} {B662}},\
  \bibinfo {pages} {53} (\bibinfo {year} {2008})},\ \Eprint
  {http://arxiv.org/abs/0711.4866} {arXiv:0711.4866 [hep-ph]} \BibitemShut
  {NoStop}%
\bibitem [{\citenamefont {Bjorken}\ \emph {et~al.}(2009)\citenamefont
  {Bjorken}, \citenamefont {Essig}, \citenamefont {Schuster},\ and\
  \citenamefont {Toro}}]{Bjorken:2009mm}%
  \BibitemOpen
  \bibfield  {author} {\bibinfo {author} {\bibfnamefont {J.~D.}\ \bibnamefont
  {Bjorken}}, \bibinfo {author} {\bibfnamefont {R.}~\bibnamefont {Essig}},
  \bibinfo {author} {\bibfnamefont {P.}~\bibnamefont {Schuster}}, \ and\
  \bibinfo {author} {\bibfnamefont {N.}~\bibnamefont {Toro}},\ }\href {\doibase
  10.1103/PhysRevD.80.075018} {\bibfield  {journal} {\bibinfo  {journal} {Phys.
  Rev.}\ }\textbf {\bibinfo {volume} {D80}},\ \bibinfo {pages} {075018}
  (\bibinfo {year} {2009})},\ \Eprint {http://arxiv.org/abs/0906.0580}
  {arXiv:0906.0580 [hep-ph]} \BibitemShut {NoStop}%
\bibitem [{\citenamefont {Ilten}\ \emph {et~al.}(2018)\citenamefont {Ilten},
  \citenamefont {Soreq}, \citenamefont {Williams},\ and\ \citenamefont
  {Xue}}]{Ilten:2018crw}%
  \BibitemOpen
  \bibfield  {author} {\bibinfo {author} {\bibfnamefont {P.}~\bibnamefont
  {Ilten}}, \bibinfo {author} {\bibfnamefont {Y.}~\bibnamefont {Soreq}},
  \bibinfo {author} {\bibfnamefont {M.}~\bibnamefont {Williams}}, \ and\
  \bibinfo {author} {\bibfnamefont {W.}~\bibnamefont {Xue}},\ }\href {\doibase
  10.1007/JHEP06(2018)004} {\bibfield  {journal} {\bibinfo  {journal} {J. High
  Energy Phys.}\ }\textbf {\bibinfo {volume} {06}},\ \bibinfo {pages} {004}
  (\bibinfo {year} {2018})},\ \Eprint {http://arxiv.org/abs/1801.04847}
  {arXiv:1801.04847 [hep-ph]} \BibitemShut {NoStop}%
\bibitem [{\citenamefont {Holdom}(1986)}]{Holdom:1985ag}%
  \BibitemOpen
  \bibfield  {author} {\bibinfo {author} {\bibfnamefont {B.}~\bibnamefont
  {Holdom}},\ }\href {\doibase 10.1016/0370-2693(86)91377-8} {\bibfield
  {journal} {\bibinfo  {journal} {Phys. Lett.}\ }\textbf {\bibinfo {volume}
  {166B}},\ \bibinfo {pages} {196} (\bibinfo {year} {1986})}\BibitemShut
  {NoStop}%
\bibitem [{\citenamefont {Redondo}(2015)}]{Redondo:2015iea}%
  \BibitemOpen
  \bibfield  {author} {\bibinfo {author} {\bibfnamefont {J.}~\bibnamefont
  {Redondo}},\ }\href {\doibase 10.1088/1475-7516/2015/07/024} {\bibfield
  {journal} {\bibinfo  {journal} {J. Cosmol. Astropart. Phys.}\ }\textbf
  {\bibinfo {volume} {1507}},\ \bibinfo {pages} {024} (\bibinfo {year}
  {2015})},\ \Eprint {http://arxiv.org/abs/1501.07292} {arXiv:1501.07292
  [hep-ph]} \BibitemShut {NoStop}%
\bibitem [{\citenamefont {Redondo}(2008)}]{Redondo:2008aa}%
  \BibitemOpen
  \bibfield  {author} {\bibinfo {author} {\bibfnamefont {J.}~\bibnamefont
  {Redondo}},\ }\href {\doibase 10.1088/1475-7516/2008/07/008} {\bibfield
  {journal} {\bibinfo  {journal} {J. Cosmol. Astropart. Phys.}\ }\textbf
  {\bibinfo {volume} {0807}},\ \bibinfo {pages} {008} (\bibinfo {year}
  {2008})},\ \Eprint {http://arxiv.org/abs/0801.1527} {arXiv:0801.1527
  [hep-ph]} \BibitemShut {NoStop}%
\bibitem [{\citenamefont {Redondo}\ and\ \citenamefont
  {Raffelt}(2013)}]{Redondo:2013lna}%
  \BibitemOpen
  \bibfield  {author} {\bibinfo {author} {\bibfnamefont {J.}~\bibnamefont
  {Redondo}}\ and\ \bibinfo {author} {\bibfnamefont {G.}~\bibnamefont
  {Raffelt}},\ }\href {\doibase 10.1088/1475-7516/2013/08/034} {\bibfield
  {journal} {\bibinfo  {journal} {J. Cosmol. Astropart. Phys.}\ }\textbf
  {\bibinfo {volume} {1308}},\ \bibinfo {pages} {034} (\bibinfo {year}
  {2013})},\ \Eprint {http://arxiv.org/abs/1305.2920} {arXiv:1305.2920
  [hep-ph]} \BibitemShut {NoStop}%
\bibitem [{\citenamefont {Braaten}\ and\ \citenamefont
  {Segel}(1993)}]{Braaten:1993jw}%
  \BibitemOpen
  \bibfield  {author} {\bibinfo {author} {\bibfnamefont {E.}~\bibnamefont
  {Braaten}}\ and\ \bibinfo {author} {\bibfnamefont {D.}~\bibnamefont
  {Segel}},\ }\href {\doibase 10.1103/PhysRevD.48.1478} {\bibfield  {journal}
  {\bibinfo  {journal} {Phys. Rev.}\ }\textbf {\bibinfo {volume} {D48}},\
  \bibinfo {pages} {1478} (\bibinfo {year} {1993})},\ \Eprint
  {http://arxiv.org/abs/hep-ph/9302213} {arXiv:hep-ph/9302213 [hep-ph]}
  \BibitemShut {NoStop}%
\bibitem [{\citenamefont {Lifshitz}\ and\ \citenamefont
  {Pitaevski}(1981)}]{ll10}%
  \BibitemOpen
  \bibfield  {author} {\bibinfo {author} {\bibfnamefont {E.}~\bibnamefont
  {Lifshitz}}\ and\ \bibinfo {author} {\bibfnamefont {L.}~\bibnamefont
  {Pitaevski}},\ }\href@noop {} {\emph {\bibinfo {title} {Course of Theoretical
  Physics}}},\ Vol.~\bibinfo {volume} {10}\ (\bibinfo  {publisher} {Pergamon},\
  \bibinfo {address} {Amsterdam},\ \bibinfo {year} {1981})\BibitemShut
  {NoStop}%
\bibitem [{\citenamefont {Johansson}(1964)}]{JOHANSSON1964378}%
  \BibitemOpen
  \bibfield  {author} {\bibinfo {author} {\bibfnamefont {S.~A.}\ \bibnamefont
  {Johansson}},\ }\href {\doibase https://doi.org/10.1016/0029-5582(64)90017-3}
  {\bibfield  {journal} {\bibinfo  {journal} {Nuclear Physics}\ }\textbf
  {\bibinfo {volume} {60}},\ \bibinfo {pages} {378 } (\bibinfo {year}
  {1964})}\BibitemShut {NoStop}%
\bibitem [{\citenamefont {Albinsson}(1971)}]{1402-4896-3-3-4-005}%
  \BibitemOpen
  \bibfield  {author} {\bibinfo {author} {\bibfnamefont {H.}~\bibnamefont
  {Albinsson}},\ }\href {http://iopscience.iop.org/1402-4896/3/3-4/005}
  {\bibfield  {journal} {\bibinfo  {journal} {Physica Scripta}\ }\textbf
  {\bibinfo {volume} {3}},\ \bibinfo {pages} {113} (\bibinfo {year}
  {1971})}\BibitemShut {NoStop}%
\bibitem [{\citenamefont {Park}(2017)}]{Park:2017prx}%
  \BibitemOpen
  \bibfield  {author} {\bibinfo {author} {\bibfnamefont {H.}~\bibnamefont
  {Park}},\ }\href {\doibase 10.1103/PhysRevLett.119.081801} {\bibfield
  {journal} {\bibinfo  {journal} {Phys. Rev. Lett.}\ }\textbf {\bibinfo
  {volume} {119}},\ \bibinfo {pages} {081801} (\bibinfo {year} {2017})},\
  \Eprint {http://arxiv.org/abs/1705.02470} {arXiv:1705.02470 [hep-ph]}
  \BibitemShut {NoStop}%
\bibitem [{\citenamefont {An}\ \emph {et~al.}(2013)\citenamefont {An},
  \citenamefont {Pospelov},\ and\ \citenamefont {Pradler}}]{An:2013yfc}%
  \BibitemOpen
  \bibfield  {author} {\bibinfo {author} {\bibfnamefont {H.}~\bibnamefont
  {An}}, \bibinfo {author} {\bibfnamefont {M.}~\bibnamefont {Pospelov}}, \ and\
  \bibinfo {author} {\bibfnamefont {J.}~\bibnamefont {Pradler}},\ }\href
  {\doibase 10.1016/j.physletb.2013.07.008} {\bibfield  {journal} {\bibinfo
  {journal} {Phys. Lett.}\ }\textbf {\bibinfo {volume} {B725}},\ \bibinfo
  {pages} {190} (\bibinfo {year} {2013})},\ \Eprint
  {http://arxiv.org/abs/1302.3884} {arXiv:1302.3884 [hep-ph]} \BibitemShut
  {NoStop}%
\bibitem [{\citenamefont {Bechteler}\ \emph {et~al.}(1984)\citenamefont
  {Bechteler} \emph {et~al.}}]{FRJ}%
  \BibitemOpen
  \bibfield  {author} {\bibinfo {author} {\bibfnamefont {H.}~\bibnamefont
  {Bechteler}} \emph {et~al.},\ }\href@noop {} {\bibfield  {journal} {\bibinfo
  {journal} {Spezielle Berichte der Kernforschungsanlage Juelich}\ }\textbf
  {\bibinfo {volume} {255}},\ \bibinfo {pages} {62} (\bibinfo {year}
  {1984})}\BibitemShut {NoStop}%
\bibitem [{\citenamefont {Deniz}\ \emph {et~al.}(2010)\citenamefont {Deniz}
  \emph {et~al.}}]{Deniz:2009mu}%
  \BibitemOpen
  \bibfield  {author} {\bibinfo {author} {\bibfnamefont {M.}~\bibnamefont
  {Deniz}} \emph {et~al.} (\bibinfo {collaboration} {TEXONO}),\ }\href
  {\doibase 10.1103/PhysRevD.81.072001} {\bibfield  {journal} {\bibinfo
  {journal} {Phys. Rev.}\ }\textbf {\bibinfo {volume} {D81}},\ \bibinfo {pages}
  {072001} (\bibinfo {year} {2010})},\ \Eprint {http://arxiv.org/abs/0911.1597}
  {arXiv:0911.1597 [hep-ex]} \BibitemShut {NoStop}%
\bibitem [{\citenamefont {Lindner}\ \emph {et~al.}(2018)\citenamefont
  {Lindner}, \citenamefont {Queiroz}, \citenamefont {Rodejohann},\ and\
  \citenamefont {Xu}}]{Lindner:2018kjo}%
  \BibitemOpen
  \bibfield  {author} {\bibinfo {author} {\bibfnamefont {M.}~\bibnamefont
  {Lindner}}, \bibinfo {author} {\bibfnamefont {F.~S.}\ \bibnamefont
  {Queiroz}}, \bibinfo {author} {\bibfnamefont {W.}~\bibnamefont {Rodejohann}},
  \ and\ \bibinfo {author} {\bibfnamefont {X.-J.}\ \bibnamefont {Xu}},\ }\href
  {\doibase 10.1007/JHEP05(2018)098} {\bibfield  {journal} {\bibinfo  {journal}
  {J. High Energy Phys.}\ }\textbf {\bibinfo {volume} {05}},\ \bibinfo {pages}
  {098} (\bibinfo {year} {2018})},\ \Eprint {http://arxiv.org/abs/1803.00060}
  {arXiv:1803.00060 [hep-ph]} \BibitemShut {NoStop}%
\bibitem [{\citenamefont {Jaeckel}\ and\ \citenamefont
  {Ringwald}(2010)}]{Jaeckel:2010ni}%
  \BibitemOpen
  \bibfield  {author} {\bibinfo {author} {\bibfnamefont {J.}~\bibnamefont
  {Jaeckel}}\ and\ \bibinfo {author} {\bibfnamefont {A.}~\bibnamefont
  {Ringwald}},\ }\href {\doibase 10.1146/annurev.nucl.012809.104433} {\bibfield
   {journal} {\bibinfo  {journal} {Ann. Rev. Nucl. Part. Sci.}\ }\textbf
  {\bibinfo {volume} {60}},\ \bibinfo {pages} {405} (\bibinfo {year} {2010})},\
  \Eprint {http://arxiv.org/abs/1002.0329} {arXiv:1002.0329 [hep-ph]}
  \BibitemShut {NoStop}%
\bibitem [{\citenamefont {Ko}\ \emph {et~al.}(2017)\citenamefont {Ko} \emph
  {et~al.}}]{Ko:2016owz}%
  \BibitemOpen
  \bibfield  {author} {\bibinfo {author} {\bibfnamefont {Y.}~\bibnamefont {Ko}}
  \emph {et~al.},\ }\href {\doibase 10.1103/PhysRevLett.118.121802} {\bibfield
  {journal} {\bibinfo  {journal} {Phys. Rev. Lett.}\ }\textbf {\bibinfo
  {volume} {118}},\ \bibinfo {pages} {121802} (\bibinfo {year} {2017})},\
  \Eprint {http://arxiv.org/abs/1610.05134} {arXiv:1610.05134 [hep-ex]}
  \BibitemShut {NoStop}%
\end{thebibliography}%

\end{document}